\documentstyle[12pt]{article}

\newcommand{\EQ}{\begin{equation}}
\newcommand{\EN}{\end{equation}}
\newcommand{\bea}{\begin{eqnarray}}
\newcommand{\ena}{\end{eqnarray}}
\newcommand{\vs}[1]{\vspace{#1 mm}}

\renewcommand{\b}{\beta}
\renewcommand{\c}{\gamma}

\renewcommand{\t}{\theta}
\newcommand{\tb}{{\bar \theta}}
\newcommand{\shalf}{\frac{1}{2}}
\newcommand{\pa}{\partial}
\newcommand{\dz}{\frac{dzd^2\t}{2\pi i}}
\newcommand{\nn}{\nonumber \\}

\begin{document}

\topmargin 0pt
\oddsidemargin 5mm

\renewcommand{\Im}{{\rm Im}\,}
\newcommand{\NP}[1]{Nucl.\ Phys.\ {\bf #1}}
\newcommand{\PL}[1]{Phys.\ Lett.\ {\bf #1}}
\newcommand{\CMP}[1]{Comm.\ Math.\ Phys.\ {\bf #1}}
\newcommand{\PR}[1]{Phys.\ Rev.\ {\bf #1}}
\newcommand{\PRL}[1]{Phys.\ Rev.\ Lett.\ {\bf #1}}
\newcommand{\PTP}[1]{Prog.\ Theor.\ Phys.\ {\bf #1}}
\newcommand{\PTPS}[1]{Prog.\ Theor.\ Phys.\ Suppl.\ {\bf #1}}
\newcommand{\MPL}[1]{Mod.\ Phys.\ Lett.\ {\bf #1}}
\newcommand{\IJMP}[1]{Int.\ Jour.\ Mod.\ Phys.\ {\bf #1}}

\begin{titlepage}
\setcounter{page}{0}
\begin{flushright}
NBI-HE-94-08\\
February 1994\\
hep-th/9402042
\end{flushright}

\vs{8}
\begin{center}
{\Large TOWARD THE UNIVERSAL THEORY OF STRINGS}

\vs{15}
{\large Fiorenzo Bastianelli\footnote{e-mail address:
 fiorenzo@nbivax.nbi.dk}}\\
{\em The Niels Bohr Institute, Blegdamsvej 17, DK-2100 Copenhagen \O,
Denmark}

\vs{8}
{\large Nobuyoshi Ohta\footnote{e-mail address:
ohta@nbivax.nbi.dk, ohta@fuji.wani.osaka-u.ac.jp}$^,$\footnote{
Permanent address: Institute of Physics, College of General Education,
Osaka University, Toyonaka, Osaka 560, Japan}}\\
{\em NORDITA, Blegdamsvej 17, DK-2100 Copenhagen \O, Denmark}\\

\vs{8}
{\large Jens Lyng Petersen\footnote{e-mail address:
 jenslyng@nbivax.nbi.dk}}\\
{\em The Niels Bohr Institute, Blegdamsvej 17, DK-2100 Copenhagen \O,
Denmark}
\end{center}

\vs{8}
\centerline{{\bf{Abstract}}}

We show that the $N=2$ superstrings may be viewed as a special class of
the $N=4$ superstrings and demonstrate their equivalence. This allows us to
realize all known string theories based on linear algebras and with $N<4$
supersymmetries as special choices of the vacua in the $N=4$ superstring.

\end{titlepage}
\newpage
\renewcommand{\thefootnote}{\arabic{footnote}}
\setcounter{footnote}{0}

One of the fascinating features of string theories is their high
degree of uniqueness. The choice of their vacua seems to be the only
freedom we have. It is therefore tempting to contemplate that there may
be a universal theory from which all string theories are derived
just by selecting different vacua. The present formulation of string
theories, however, is too rigid to make the freedom of choosing
a vacuum manifest.

A recent discovery by Berkovits and Vafa~\cite{BV} has opened a way to
interpret all string theories as kinds of spontaneously broken phases of
string theories with higher world-sheet symmetries. In particular, they
have shown that the $N=0~(N=1)$ strings can be viewed as a special class
of vacua for the $N=1~(N=2)$ superstrings. The equivalence of these $N=1
{}~(N=2)$ and $N=0~(N=1)$ strings has been discussed in the original
paper~\cite{BV}, where it has been shown that the scattering amplitudes
coincide, and this is further confirmed in refs.~\cite{FO,IK,FB,OP}. Given
these results, it is natural to speculate that $N=2$ string theories
can also be embedded into $N=4$ string theories~\cite{N4S}. The main obstacle
in this subject at the moment seems to be that nobody knows how to compute the
scattering (loop) amplitudes for general $N=4$ string theories.
However, we can use the method in refs.~\cite{IK,OP} which makes use
of a similarity transformation to relate the two theories. This
will enable us to show the equivalence for scattering amplitudes
without going into the details of how to compute them, and will give an
isomorphism between the operator algebras.

The purpose of this paper is to realize the $N=2$ superstring as a special
choice of the vacua in the $N=4$ superstring, and to give a simple and
explicit proof that our $N=4$ formulation is equivalent to the $N=2$
superstring.
This completes the program of embedding string theories with $N<4$
supersymmetries into a universal string theory, which turns out to be
the so-called `large' $N=4$ superstring with $O(4)$ symmetry~\cite{N4S,STV}.

In order to embed the $N=2$ superstrings with matter central charge
$c_m=6$ into $N=4$, it might be natural to first try the `small' $N=4$ string
with $SU(2)$ symmetry~\cite{N4S}. Since the ghosts for this theory
consist of one fermionic $(b,c)$, four bosonic $(\b,\c)$ and three
fermionic $(b_1,c_1)$ systems with spins $(2,1),(\frac{3}{2},-\shalf)$
and $(1,0)$, respectively, it would be necessary to introduce two fermionic
$(\eta,\xi)$ and two bosonic $(\b,\c)$ systems with spins $(\frac{3}{2},
-\shalf)$ and $(1,0)$ as additional matter fields in order to cancel
the extra ghosts compared with the $N=2$
superstring. These additional matter fields carry the right central
charge $-18$, so that one will get a system with $c=-12$,
appropriate for the matter system of the `small' $N=4$ superstring.
{}From previous experiences in the case of lower supersymmetry~\cite{BV},
one would then expect that one of the additional supersymmetry generators
may be obtained just by improving the $N=2$ BRST current by total
derivative terms such that it becomes nilpotent. It turns out that one
can indeed uniquely improve the BRST current this way. However,
the whole algebra that is satisfied by all the generators is not the
small one but the so-called large $N=4$ superconformal algebra with
zero central charge~\cite{GR}.

We will use $N=2$ superfields to describe the $N=4$ superstring theory as
in ref.~\cite{GR}.
Let us take the $N=2$ superstrings with matter super stress-energy tensor
$T_m$ with central charge $c_m= 6$, and add to the system two fermionic
superfields $(\eta,\xi)$, similar to the ghosts in the $N=2$ string
but with the spin shifted to the value $(\shalf,-\shalf)$. These additional
fields contribute $-6$ to the central charge,\footnote{Note that the
contribution of $N=2$ super $(b,c)$ systems to the central charge is always
$\pm 6$, independently of their spins (the sign depends on
the statistics).} resulting in a total central charge $c=0$.

The large $N=4$ superconformal algebra consists of four $N=2$ superfield
generators $T, G, {\bar G}$ and $J$, and their operator products
(OPE) are given by
\bea
T(Z_1) T(Z_2) &\sim& \frac{c/3 + \t_{12}\tb_{12} T}{z_{12}^2}
  +\frac{-\t_{12}DT + \tb_{12}{\bar D}T + \t_{12}\tb_{12}\pa T}{z_{12}}, \nn
T(Z_1){\cal O}(Z_2) &\sim& h_{\cal O}\frac{\t_{12}\tb_{12}}{z_{12}^2}{\cal O}
      + \frac{-\t_{12}D{\cal O} + \tb_{12}{\bar D}{\cal O} + \t_{12}\tb_{12}
      \pa{\cal O}}{z_{12}}, \nn
G(Z_1) {\bar G}(Z_2) &\sim& \frac{-\t_{12}DJ + \tb_{12}{\bar D}J
      - \t_{12}\tb_{12}(T- \shalf \pa J)}{z_{12}}, \nn
J(Z_1) G(Z_2) &\sim& - \frac{\t_{12}\tb_{12}}{z_{12}} G, \nn
J(Z_1) {\bar G}(Z_2) &\sim&  \frac{\t_{12}\tb_{12}}{z_{12}} {\bar G},
\ena
where $Z \equiv (z,\t,\tb)$ denotes the super coordinates, $D = \pa_\t -
\shalf \tb\pa_z$ and ${\bar D} = \pa_\tb -\shalf \t\pa_z$
are the supercovariant derivatives in $N=2$ superspace, $z_{12} \equiv
z_1 - z_2 +\shalf(\t_1\tb_2 +\tb_1\t_2), \t_{12}\equiv \t_1 -\t_2,
\tb_{12}\equiv \tb_1 -\tb_2$, and ${\cal O}$ stands for the generators
$G, {\bar G}$ and $J$ with dimensions given by $h_{\cal O}= 1/2$
for $G,{\bar G}$ and $h_{\cal O}=0$ for $J$. In general the large $N=4$
superconformal algebra~\cite{STV} has more structure with two parameters
$k^+$ and $k^-$ characterizing the levels of the two $SU(2)$ current
algebras and with central charge $c= 6k^+ k^-/(k^+ +k^-)$.
In eq.~(1), we have given the algebra for the special case of $k^+=k^-$ and
in the limit $k^+\to 0$ which is the one realized in the $N=2$
theory.\footnote{The structure of the general $N=4$
superconformal symmetry in terms of components and $N=2$ superfields and its
BRST operator will be discussed elsewhere~\cite{BO}.}

The matter generators for the $N=4$ algebra turn out to be given by~\cite{GR}
\bea
T &=& T_m - \shalf \pa(\eta\xi) + (D\xi)({\bar D}\eta) + ({\bar D}\xi)(D\eta),
\nn
G &=& \xi T_m + \xi(D\xi)({\bar D}\eta) + \xi({\bar D}\xi)(D\eta)
      - \eta(D\xi)({\bar D}\xi), \nn
{\bar G} &=& - \eta, \nn
J &=& \eta \xi.
\ena
Using the OPE for $T_m$ which is the same as that given in eq.~(1) with
$c=6$ and the correlation
\EQ
\xi(Z_1)\eta(Z_2) \sim \frac{\t_{12}\tb_{12}}{z_{12}},
\EN
it is easy to check
that these indeed satisfy the OPE (1) with central charge $c=0$.

The BRST operator for this $N=4$ superstring takes the form
\bea
Q_{N=4}&=& \oint\dz \left[ C_t \left( T + \shalf T_{gh} \right)
      + C_g \left( G + \shalf G_{gh} \right) \right. \nn
&+& \left. C_{\bar g} \left({\bar G} + \shalf{\bar G}_{gh} \right)
      + C_j \left( J + \shalf J_{gh} \right) \right],
\ena
where the four sets of fields $(C_t,B_t),(C_g,B_g),(C_{\bar g},B_{\bar g})$
and $(C_j,B_j)$ with spins $(-1,1),(-\shalf,\shalf),$ $(-\shalf,\shalf)$ and
$(0,0)$ are the reparametrization, supersymmetry and current
ghosts in $N=2$ superfields for the large $N=4$ superstrings with the
correlations
\EQ
C(Z_1) B(Z_2) \sim \frac{\t_{12}\tb_{12}}{z_{12}}.
\EN
In eq.~(4), the generators with subscript $gh$ are those for ghosts:
\bea
T_{gh} &=& - \pa(B_t C_t) + ({\bar D}C_t)(DB_t) + (DC_t)({\bar D}B_t) \nn
&-& \shalf\pa(B_g C_g) + ({\bar D}C_g)(DB_g) + (DC_g)({\bar D}B_g) \nn
&-& \shalf\pa(B_{\bar g} C_{\bar g}) + ({\bar D}C_{\bar g})(DB_{\bar g})
 +(DC_{\bar g})({\bar D}B_{\bar g}) \nn
&+& \ ({\bar D}C_j)(DB_j) + (DC_j)({\bar D}B_j), \nn
G_{gh} &=& \shalf C_{\bar g} \pa B_j + C_t \pa B_g + C_{\bar g}B_t
      + ({\bar D}C_t)(DB_g) + (DC_t)({\bar D}B_g) \nn
&-& C_j B_g - ({\bar D}C_{\bar g})(DB_j)
  - (DC_{\bar g})({\bar D}B_j) + \shalf \pa C_t B_g, \nn
{\bar G}_{gh} &=& -\shalf C_g \pa B_j + C_t \pa B_{\bar g} + C_g B_t
      + (DC_t)({\bar D}B_{\bar g}) + ({\bar D}C_t)(DB_{\bar g}) \nn
&+& C_j B_{\bar g} + (DC_g)({\bar D}B_j)
  + ({\bar D}C_g)(DB_j) + \shalf \pa C_t B_{\bar g}, \nn
J_{gh} &=& C_t \pa B_j - C_g B_g + C_{\bar g}B_{\bar g}
 + (DC_t)({\bar D} B_j) + ({\bar D}C_t)(DB_j).
\ena
It is easy to check that these generators also satisfy the OPE in eq.~(1)
with $c=0$.

Substituting (2) and (6) into (4), we find after partial integration
\bea
Q_{N=4} &=& \oint \dz \left[ \left(C_t +C_g\xi \right) T_m
 + C_t \left( \shalf \pa(\xi\eta) + (D\xi)({\bar D}\eta)
 + ({\bar D}\xi)(D\eta) \right) \right. \nn
&+& C_g \left( \xi(D\xi)({\bar D}\eta) + \xi({\bar D}\xi)(D\eta)
 - \eta ({\bar D}\xi)(D\xi) \right) - C_{\bar g}\eta - \xi\eta C_j \nn
&+& C_t \left( \shalf \pa C_tB_t + \shalf ({\bar D}C_t)(DB_t)
 + \shalf (DC_t)({\bar D}B_t) - \frac{1}{2} \pa (C_gB_g) \right.\nn
&+& ({\bar D}C_g)(DB_g) + (DC_g)({\bar D}B_g)
 - \frac{1}{2} \pa (C_{\bar g}B_{\bar g})
 +  ({\bar D}C_{\bar g})( DB_{\bar g}) \nn
&+& \left. (DC_{\bar g})({\bar D}B_{\bar g})
 +  ({\bar D}C_j)(DB_j) +  ( DC_j)({\bar D}B_j) \right) \nn
&+&  C_j(C_{\bar g} B_{\bar g} - C_gB_g)  + C_g C_{\bar g}B_t \nn
&+& \left. C_{\bar g} \left( - \shalf C_g \pa B_j +
({\bar D}C_g)(DB_j) +  (DC_g)({\bar D}B_j)  \right)
\right].
\ena
It is straightforward but tedious to check the nilpotency of this BRST operator
directly since the OPE of the BRST current with itself produces quite a large
number of terms
which add up to total derivatives. The equivalent result can be proved
if the double commutators of the BRST charge with all the fundamental fields
vanish since this means that the square of the charge vanishes in the Hilbert
space of the theory. By this method as well as the direct method, we have
checked that this BRST charge has the desired properties, and in particular
that it is indeed nilpotent.

We would like to compare this BRST operator with that of the $N=2$ superstring:
\EQ
Q_{N=2} = \oint\dz C_t \left[ T_m + \shalf\pa C_t B_t
 + \shalf(DC_t)({\bar D}B_t) + \shalf({\bar D}C_t)(DB_t) \right].
\EN

Our first step to show the equivalence is to transform the BRST operator
(7) into a direct sum of that for the $N=2$ superstrings (8) with some
additional terms and those for topological sectors by a similarity
transformation
\EQ
e^{R} Q_{N=4} e^{-R} = {\tilde Q}_{N=2} + Q_{top} + Q_{U(1)},
\EN
where
\bea
R &=& \oint\dz \xi \left[ - C_g B_t - C_j B_{\bar g}
 - \shalf B_{\bar g}\pa C_t - \pa B_{\bar g} C_t - ({\bar D}B_{\bar g})(DC_t)
 \right.\nn
&-& \left. (DB_{\bar g})({\bar D}C_t)
 - ({\bar D}C_g)(DB_j) + \shalf C_g\pa B_j -(DC_g)({\bar D}B_j) \right], \\
{\tilde Q}_{N=2} &=& Q_{N=2} + \oint\dz C_t \left[
 - \shalf\pa(C_g B_g) + (DC_g)({\bar D} B_g)
\right. \nn
&+& \left. ({\bar D} C_g)(DB_g)
 + (DC_j)({\bar D} B_j) + ({\bar D} C_j)(DB_j) \right],
\ena
and \footnote{Note that the additional terms in ${\tilde Q}_{N=2}$ are
just the products of $C_t$ and the free super-stress tensors for
the ghosts $C_g,B_g,C_j$ and $B_j$.} where
\bea
Q_{top} &=& -\oint\dz C_{\bar g}\eta, \nn
Q_{U(1)} &=& -\oint\dz C_g B_g C_j,
\ena
are the BRST operators for the topological sectors. It is easy to confirm
that the operators on the right hand side of (9) all anticommute
with one another and are nilpotent. From eq. (9), we see that the cohomology
of the $Q_{N=4}$ is a direct product of those of ${\tilde Q}_{N=2}, Q_{top}$
and $Q_{U(1)}$.

The BRST operator $Q_{top}$ imposes the condition that
$B_{\bar g}, C_{\bar g}, \eta$ and $\xi$ fall into the quartet representations
of the BRST operator and all decouple from the physical subspace. Consequently
the cohomology of the $Q_{top}$ consists only of their vacuum.

The condition imposed by $Q_{U(1)}$ tells us that the modes $B_g ,C_g,B_j$
and $C_j$ also make quartets and decouple from the physical subspace
except for the combinations $C_gB_g$ and its arbitrary powers all of which
have vanishing ghost number. However, those combinations with zero ghost
number have zero norm because the OPE of the current $C_gB_g$ with itself
is non-singular. It follows that these modes also completely decouple from
the physical subspace.
This means that we can disregard the additional terms in ${\tilde Q}_{N=2}$
which then reduces to the ordinary $N=2$ BRST operator $Q_{N=2}$.

Thus the cohomology of our theory is basically reduced to that of the $N=2$
superstring and we obtain a one-to-one
correspondence between the cohomologies of $Q_{N=4}$ and $Q_{N=2}$.
It is also clear that the transformation manifestly preserves the operator
algebra. This establishes the equivalence of the $N=2$ superstrings and
the special classes of $N=4$ superstrings described above.\footnote{
It seems possible to introduce additional fermionic $(\eta_g,\xi_g)$ and
bosonic $(\eta_j,\xi_j)$ matter into the theory and cancel the remaining
free ghosts $(B_g,C_g),(B_j,C_j)$ from the theory. Specifically we find
that the following generators constitute an $N=4$ algebra with $c=0$:
\begin{eqnarray*}
T &=& \shalf \pa(\xi_g\eta_g) + (D\xi_g)({\bar D}\eta_g)
+ ({\bar D}\xi_g)(D\eta_g)+ (D\xi_j)({\bar D}\eta_j)
+ ({\bar D}\xi_j)(D\eta_j), \\
G &=& -\eta_g, \\
{\bar G} &=& \xi_g \left[ (D\xi_g)({\bar D}\eta_g)+({\bar D}\xi_g)(D\eta_g)
 + (D\xi_j)({\bar D}\eta_j)+({\bar D}\xi_j)(D\eta_j)+\shalf\pa\eta_j
\right] \\
&& - ({\bar D}\xi_g)(D\xi_g)\eta_g + (D\xi_g)({\bar D}\eta_j)
+ ({\bar D}\eta_g)(D\eta_j), \\
J &=& -\eta_j + \xi_g\eta_g.
\end{eqnarray*}
We believe that there also exists a similarity transformation to bring
the whole BRST charge into a sum of $Q_{N=2}$ and simple topological terms,
thereby simplifying our proof. This reformulation is analogous to the
formulation of QED with free Faddeev-Popov ghosts whereas ours corresponds
to the Gupta-Bleuler formalism.}

We have thus succeeded in embedding the $N=2$ superstring into the $N=4$
superstring in such a way that it is completely equivalent to the $N=2$ string.
Combined with the existing embeddings of strings into those with higher
symmetries, we now have a universal string theory realized as the $N=4$
superstring, from which all the known string theories with $N<4$
supersymmetries may be obtained by particular choices of vacua.

It remains to be seen if we can also get the small $N=4$ superstring with
$SU(2)$ symmetry from this theory. This investigation also suggests
that if one can find an improvement of the BRST
current by total derivatives such that its square is zero, one can further
extend the theory to higher string theory.
Another line of investigation is to try to embed strings into $W$-strings.
An embedding of the bosonic string into the $W_3$ string has been found in
ref.~\cite{BFW}. It would be interesting to examine these issues in the
light of our approach.

\vs{10}
\noindent
{\it Acknowledgements}

N. O. would like to thank Paolo Di Vecchia for discussions,
support and kind hospitality at NORDITA, where this work was done.
J. L. P. acknowledges useful advice on the use of the $N=2$
programme from J. O. Madsen.
Many of the calculations in this paper were performed by using the OPE
package developed by Kris Thielemans and Sergey Krivonos, whose software
is gratefully acknowledged.

\end{document}